\documentclass[aps,prb,showpacs,superscriptaddress,twocolumn]{revtex4}
\usepackage{amsmath}
\usepackage{amsfonts}
\usepackage{graphicx}

\begin{document}

\title{Dynamics of molecular nanomagnets in time-dependent external magnetic
fields: \\
Beyond the Landau-Zener-St\"{u}ckelberg model}
\author{P. F\"{o}ldi}
\affiliation{Department of Theoretical Physics, University of Szeged, Tisza Lajos k\"{o}r%
\'{u}t 84, H-6720 Szeged, Hungary}
\affiliation{Departement Fysica,
Universiteit Antwerpen, Groenenborgerlaan 171, B-2020 Antwerpen, Belgium}
\author{M. G. Benedict}
\affiliation{Department of Theoretical Physics, University of Szeged, Tisza Lajos k\"{o}r%
\'{u}t 84, H-6720 Szeged, Hungary}
\author{J. M. Pereira}
\affiliation{Departement Fysica, Universiteit Antwerpen, Groenenborgerlaan
171, B-2020 Antwerpen, Belgium}
\author{F. M. Peeters}
\affiliation{Departement Fysica, Universiteit Antwerpen, Groenenborgerlaan
171, B-2020 Antwerpen, Belgium}

\begin{abstract}
The time evolution of the magnetization of a magnetic molecular crystal is
obtained in an external time-dependent magnetic field, with sweep rates in the
kT/s range. We present the 'exact numerical' solution of the time dependent
Schr\"{o}dinger equation, and show that the steps in the hysteresis curve can
be described as a sequence of two-level transitions between adiabatic states.
The multilevel nature of the problem causes the transition probabilities to
deviate significantly from the predictions of the Landau-Zener-St\"{u}ckelberg
model. These calculations allow the introduction of an efficient approximation
method that accurately reproduces the exact results. When including phase
relaxation by means of an appropriate master equation, we observe an interplay
between coherent dynamics and decoherence. This decreases the size of the
magnetization steps at the transitions, but does not modify qualitatively the
physical picture obtained without relaxation.
\end{abstract}

 \maketitle

\section{Introduction}
There has been an increased interest in the study of crystals consisting of
high-spin molecules such as Mn$_{12}$-Ac and Fe$_{8}$O. These organic
molecules, also known as molecular nanomagnets,\cite{GSV06} contain transition
metal atoms with strongly exchange-coupled spins, which causes the individual
molecules to behave as a single, large spin. Experiments on the magnetization
dynamics of these molecular crystals have shown the presence of a series of
steps in the hysteresis curve at sufficiently low
temperatures.\cite{FST96,MSS01,WMC06} This behavior is a consequence of
quantum mechanical tunneling of spin states through the anisotropy energy
barrier and occurs when the external field brings two levels at different
sides of the barrier into resonance. This macroscopic quantum effect has been
the subject of intensive experimental and theoretical investigation which
revealed additional remarkable properties of the magnetic molecules. In
particular, it has been demonstrated that the magnetic tunneling could be
accompanied by the emission of electromagnetic radiation, and bursts of
microwave pulses have been detected in recent
experiments.\cite{TCHA04,VSM04,HJAGHT05} It has been proposed that the
physical mechanism responsible for this radiation is a collective quantum
effect known as superradiance.\cite{D54,SRAD,CG02,HK03,JCC04} This
interpretation has been questioned and it was argued that when one includes
the time scale of relaxation, a maser-like effect is more likely responsible
for the observations.\cite{BFP05} Furthermore, the change in magnetization can
be described in terms of avalanches, which were recently shown to propagate
through the crystal in an analogous way to that of a flame front in a
flammable chemical substance (deflagration).\cite{SSC05,HMH05} It has also
been suggested\cite{LL01} that these molecules can be used for implementing a
quantum computational algorithm.

In this work we study the dynamics of the multilevel system corresponding to
the $21$ spin states of the Mn$_{12}$-Ac molecule ($S=10$) in a time-dependent
magnetic field. An 'exact numerical' solution of the relevant time-dependent
Schr\"{o}dinger equation is obtained in the whole time interval of interest.
The results show that, although the usual qualitative picture of consecutive
two-level transitions holds, the transition probabilities deviate
significantly from the predictions of the
Landau-Zener-St\"{u}ckelberg\cite{L32,Z32,S32} (LZS) model. This deviation is
closely related to the multilevel nature of the problem: examples are given
where the single relevant parameter of the LZS model (which is essentially the
level splitting in appropriate units) is the same, but the transition
probabilities are different due to the different time dependence of the
adiabatic energy levels. An efficient approximation method based on non-LZS
two-level transitions is introduced, which is able to describe the dynamics
with high accuracy. Furthermore, relaxation effects are also included. By
considering realistic dephasing rates, it is shown that for field sweep rates
in the kT/s range, neither unitary time evolution nor relaxation dominates the
dynamics. The interplay between these two processes results in a decrease of
the transition probability at a given avoided level crossing.

The paper is organized as follows: In Sec.~\ref{modelsec} the relevant
Hamiltonian is discussed, along with its level structure, and the dynamical
equations in the adiabatic basis are introduced. Results related to the
solution of the time dependent Schr\"{o}dinger equation are presented in
Sec.~\ref{unitarysec}, and the consequences of relaxation effects are
discussed in Sec.~\ref{relaxsec}. Finally in Sec.~\ref{conclusionsec} the
results are summarized and conclusions are presented.

\section{Magnetic level structure and dynamical equations}

\label{modelsec} Experimental\cite{FST96,MSS01,M99,BGS97,H98,HEJ03} studies on
crystals of Mn$_{12}$Ac and Fe$_{8}$O suggest that the spin Hamiltonian for
these systems can be written as
\begin{equation}
H_{S}(t)=H_{0}(t)+H_{1}(t),  \label{H}
\end{equation}
where $H_{0}$ is diagonal in the eigenbasis $\{|m\rangle \}$ of the $z$
component of the spin operator, $S_{z}$:
\begin{equation}
H_{0}(t)=-DS_{z}^{2}-FS_{z}^{4}-g\mu _{B}B(t)S_{z}.  \label{H0}
\end{equation}
Here the last term in the right-hand side describes the coupling to an
external magnetic field applied along the $z$ direction, which is parallel
with the easy axis of the crystal. This external field is time-dependent, with
sweep rates on the kT/s scale.\cite{VSM04} $H_{1}$ in the Hamiltonian contains
terms\cite{MSS01,BGS97} that do not commute with $S_{z}$:
\begin{equation}
H_{1}=C(S_{+}^{4}+S_{-}^{4})+E(S_{+}^{2}+S_{-}^{2})/2+K(S_{+}+S_{-})/2.
\label{H1}
\end{equation}
In the present paper we will concentrate on Mn${}_{12}$-Ac, which can be
considered as a representative example of molecular nanomagnets. In this case
the values of the parameters in $H_{0}$ are $D/k_{B}=0.56K,$ and
$F/k_{B}=1.1\cdot 10^{-3}K$. The coefficients in $ H_{1}$, which are essential
for the determination of the transition probabilities, can be obtained by
fitting the theoretical results to experimental magnetization
curves.\cite{BFP05} In this paper we use $K=0.025  g\mu _{B}B$,
$E/k_{B}=-4.48$ $10^{-3}$ $K$, $C/k_{B}=1.7$ $10^{-5}$ $K$ unless otherwise
stated.

Considering the total Hamiltonian (\ref{H}) as the generator of the time
evolution, the corresponding time-dependent Schr\"{o}dinger equation governs
the dynamics. We can also use a density operator $\varrho$ to describe the
system and write
\begin{equation}
\frac{\partial \varrho}{\partial t}=-{i}\left[ H_{S}, \varrho \right],
\label{master1}
\end{equation}%
where $\hbar=1$. Relaxation effects can then be included through additional
terms on the right hand side of this equation, see Sec.~\ref{relaxsec} for
more details.

A direct calculation of the time-dependent solutions of Eq.~(\ref{master1})
when expanded in the $\{|m\rangle \}$ eigenbasis of the spin operator $S_{z}$
turns out to be a rather difficult problem: even for large field sweep rates
(i.e, kT/s), the saturation of the magnetization is reached in a few
milliseconds. During this time, roughly $10^{9}$ Bohr oscillations take place
due to $H_{0}$, raising demanding requirements on the accuracy of the
numerical process. An alternative and more efficient way of dealing with this
equation is based on the expansion of the time-dependent states in an
adiabatic basis, i.e., the basis of the instantaneous eigenstates of
$H_{S}(t)$:
\begin{equation}
H_{S}(t) \left| E_{n} (t)\right\rangle=E_{n}(t)\left| E_{n} (t)\right\rangle.
\label{adiab}
\end{equation}

It is convenient to label these states so that they correspond to the
eigenenergies in increasing order: $E_{0}(t)<E_{1}(t)<\ldots<E_{20}(t)$ at all
times. The energy curves $E_{n}(t)$ are thus continuous. It must be stressed
that the time dependence of these states is parametrical. In fact they are in
a one-to-one correspondence with the external field, and thus the value of the
time-dependent $B$ completely determines the states $\left|
E_{n}\right\rangle$. Next, the density operator is expanded in the
time-dependent basis determined by Eq.~(\ref{adiab}):
\begin{equation}
\varrho (t)=\sum_{nm}e^{i\int_{t_{0}}^{t} (E_{m}-E_{n})dt^{\prime}}\rho
_{nm}(t)\left\vert E_{n}(t)\right\rangle \left\langle E_{m}(t)\right\vert .
\label{rhoexpand}
\end{equation}%
Using Eq.~(\ref{master1}) to calculate the dynamics, and Eq.~(\ref{adiab}) to
obtain the time dependence of the adiabatic states, one finds that the time
evolution of the matrix $\rho $ takes the form of a von Neumann equation
\begin{equation}
\frac{\partial \rho }{\partial t}=-i\left[ \widetilde{H},\rho \right] ,
\label{neumann}
\end{equation}%
where $\widetilde{H}$ is given by
\begin{equation}
\widetilde{H}_{nm}(t)=i\left\langle E_{n}\right\vert \frac{\partial H_{S}}{
\partial t}\left\vert E_{m}\right\rangle \frac{e^{i\int_{t_{0}}^{t} (E_{n}-E_{m})dt^{\prime}}
}{E_{n}(t)-E_{m}(t)},  \label{ine}
\end{equation}
if $n\neq m$, and $\widetilde{H}_{nn}=0.$ This expression explicitly shows
that an appreciable change in the populations $\rho _{nn}$ and $\rho _{mm}$ is
expected around the avoided crossing of the levels $E_{n}$ and $E_{m}$, i.e.,
when the denominator in Eq.~(\ref{ine}) has a local minimum. A qualitatively
similar conclusion follows from a degenerate perturbation
calculation\cite{vV29,dC60,K74,G91,LL00,YP05,BFP05} around the avoided level
crossings; but note that Eq.~(\ref{ine}) is an exact result.

\begin{figure}[tbp]
\includegraphics[width=8.8 cm]{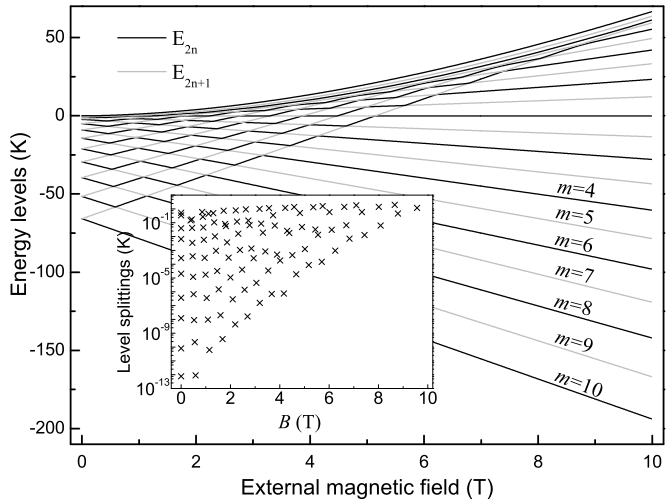}
\caption{The level scheme of the Hamiltonian (\ref{H}) as a function of the
external magnetic field $B$. The energy levels corresponding to the exact
eigenstates $\left| E_{n}\right\rangle$ have "zigzag" form, while the
approximate eigenenergies corresponding to the eigenstates of $S_{z}$ are
almost straight lines (with periodically changing black and grey sections) for
low energies. (A few examples are labeled in the figure.) The inset shows the
minimal distance between levels as a function of $B,$ the horizontal position
of the crosses coincide the avoided level crossings shown in the main part of
the figure.} \label{mindistfig}
\end{figure}

The level scheme and the minimal energy difference between levels at the
avoided crossings are shown in Fig.~\ref{mindistfig}. As a guiding line, far
from the crossings we can associate a label $m$ to each energy eigenvalue
$E_{n}$ in such a way that the overlap $\vert \langle m \left\vert
E_{n}\right\rangle \vert$ is maximal over all possible values of $m$ between
-10 and 10. This assignment is based on the fact that $H_1$ is a relatively
weak perturbation to $H_0,$ thus -- at least for low energies and far from the
crossings -- the eigenstates of the full spin Hamiltonian are close to that of
$H_{0}.$ As a consequence of the labelling convention introduced after
Eq.~(\ref{adiab}), a given adiabatic eigenstate $\left| E_{n}\right\rangle$
before and after the avoided crossing of the levels $E_{n}$ and $E_{n\pm 1}$
corresponds to two different states $ |m\rangle\neq|m^{\prime}\rangle$, see
Fig.~\ref{mindistfig}. In other words, if the population corresponding to a
certain adiabatic state does not change while passing a crossing, the
expectation value of $S_{z}$ and thus the magnetization \emph{does} change.

The fact that the level splittings at the avoided crossings can differ by 12
orders of magnitude, raises an additional difficulty when using
Eq.~(\ref{neumann}) to calculate the dynamics, because the derivatives also
change in a similarly wide range. Therefore it turned out that a combination
of Eqs.~(\ref {master1}) and (\ref{neumann}) leads to the most efficient
method: when some populated levels are too close to each other,
Eq.~(\ref{neumann}) is no longer able to provide the required accuracy,
therefore we change the basis and use Eq.~(\ref{master1}) for a short time
interval after which it will be safe again to work with Eq.~(\ref {neumann}).
Thus the control parameter defining the stepsize needed to have the required
accuracy for such a long calculation is essentially the minimal distance
between the populated levels.

The dynamical equation (\ref{neumann}) indicates that the usual approach of
treating the problem as a sequence of two-level transitions (each taking place
at the corresponding level crossing) may provide an accurate approximation to
the time evolution. In this framework the dynamics of the states corresponding
to the anticrossing levels is governed by a $2\times 2$ Hamiltonian resulting
from the reduction\cite{vV29,dC60,K74,G91,LL00,YP05,BFP05} of the complete
$H_{S}$ to the relevant level pairs. Additionally, in this approach it is
usually assumed that the time dependence of the diagonal elements of the
reduced Hamiltonian is linear, while the offdiagonal ones are constants:
\begin{equation}
H_{red}(t)=\hbar \left(
\begin{array}{cc}
\Omega t & \Delta/2 \\
\Delta/2 & -\Omega t%
\end{array}
\right).  \label{Hred}
\end{equation}
With these assumptions each avoided level crossing is identical to the
Landau-Zener-St\"{u}ckelberg\cite{L32,Z32,S32} (LZS) model, which has an
analytical solution yielding the transition probability $P_{LZS}=1-\exp (-\pi
\Delta ^{2}/2 \Omega)$ in the long-time limit. Note that small values of
$P_{LZS}$ means no appreciable change either in the population of the
eigenstates of $S_{z}$, or the magnetization (but almost complete exchange of
the populations of the adiabatic states); $P_{LZS}\approx 1$ is observable as
a step in the magnetization, while the populations of the adiabatic levels are
practically unchanged. Corrections to the LZS model originating from dipolar
interactions have been investigated in Ref.~[\onlinecite{LWF02}]. It is
important to emphasize that $P_{LZS}$ depends on the ratio
$\Delta/\sqrt{\Omega}$, i.e., on a single parameter (which, in appropriate
dimensionless units, is simply the level splitting). In the next section we
show that the dynamics in the whole spin Hilbert-space can no longer be
described by a single parameter, and consequently the exact transition
probabilities can be significantly different from $P_{LZS}.$

\section{Unitary time evolution}

\label{unitarysec} 
\begin{figure}[tbp]
\includegraphics[width=8.8 cm]{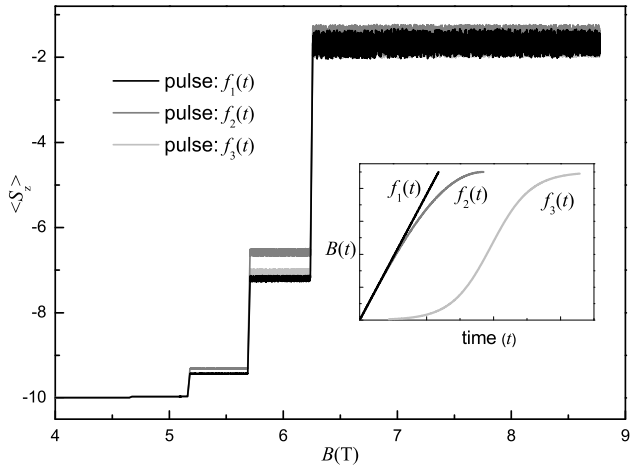}
\caption{The expectation value of $\langle S_{z}\rangle$ as a function of the
time dependent external magnetic field $B(t)$, which is shown in the inset,
for three different pulse shapes with $w=2$ kT/s.} \label{magnfig2}
\end{figure}

In this section we calculate the unitary dynamics described
by Eq.~(\ref{master1}). Initially the external magnetic field is zero, then it
raises to its maximal value of $B_{\max },$ i.e., $B(t)=f(t)B_{\max }.$ Here
we consider three analytical shapes of the function $f(t)$: a linear, a sine
and a tangent hyperbolical pulse, see the inset in Fig.~\ref{magnfig2}. We
construct these pulses in such a way, that the maximal external magnetic field
rate $ w=\max(dB/dt)=B_{max}\max(df/dt)$ is the same and falls in the kT/s
range:
\begin{subequations}
\label{pulses}
\begin{eqnarray} f_{1}(t)&=&w t, \label{pulse1}
\\ f_{2}(t)&=&\sin\left(\frac{wt}{B_{\max }}\right), \label{pulse2}
\\ f_{3}(t)&=&\frac{1}{2} \left [\tanh(\frac{2wt-\delta}{B_{\max }})+1
\right], \label{pulse3}
\end{eqnarray}
\end{subequations}
where the shift $\delta$ in $f_{3}(t)$ has to be chosen such that at $t=0$ the
external magnetic field is negligible.
\begin{figure}[tbp]
\includegraphics[width=8.8 cm]{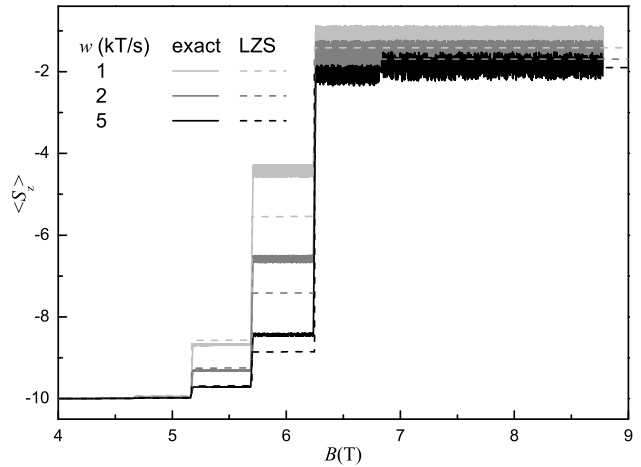}
\caption{The expectation value of $\langle S_{z}\rangle$ (solid lines) as a
function of the external magnetic field $B$ for different sweep rates. The
exact results are compared with the predictions of the LZS approximation
(dashed lines). The pulse shape corresponding to this figure is given by
Eq.~(\ref{pulse2}). Above 5.7 T rapid oscillations appear, see the text for
more details.} \label{magnfig1}
\end{figure}

The initial state at the beginning of the calculation ($t=0 $) is the lowest
energy eigenstate that later crosses other adiabatic states, i.e., $\left\vert
\Psi \right\rangle (0)=\left\vert E_{1}\right\rangle \approx |m=-10\rangle.$
This means that we follow the lowest increasing curve on the level scheme
shown in Fig.~\ref{mindistfig}, and the energy levels that cross this line
correspond to decreasing energies and thus do not meet any other levels later.
Similarly, if initially the ground state $\vert E_{0}\rangle$ was populated,
it would not give any contribution to the steps in the magnetization curve, it
simply leads, to a very good approximation, to an additional constant. (Which
is the reason for our choice of the initial state.)

The expectation value $\langle S_{z}\rangle $ as a function of the external
magnetic field $B$ is shown in Fig.~\ref{magnfig1} for the sine pulse
(\ref{pulse2}) and different values of the maximal sweep rate $w.$ The steps
seen in this figure are very similar to the experimental curves, but differ
from the result that can be obtained by using the LZS theory (also plotted in
Fig.~\ref{magnfig1}). Faster sweep rates mean smaller transition probabilities
between the eigenstates of $S_{z}$. Although the exact dynamics is different
from the LZS result, in the investigated sweep rate range we found that
$\langle S_{z}\rangle$ scales with the sweep rate almost exactly the same way
as one could deduce from $P_{LZS}$. Additionally, Fig.~\ref{magnfig1} also
shows that since the states $\vert m\rangle$ are not exact eigenstates of the
complete spin Hamiltonian $H_{S}$, there are rapid oscillations in $\langle
S_{z}\rangle$ for higher external fields, which are clear indications of the
Bohr oscillations corresponding to different eigenenergies of $H_{S}.$ We will
find that if we take relaxation effects into account (Sec.~\ref{relaxsec}),
these oscillations disappear on a very short timescale.

Comparison between the results for $\langle S_{z} \rangle$ for different
functional shapes of the external magnetic field $B(t)$ is shown in
Fig.~\ref{magnfig2} in case of a maximal sweep rate of $w=2$ kT/s. Note that
the difference is not too large; in this sweep rate range it is not the
functional time-dependence of $B(t)$ that determines the heights of the steps
seen in the magnetization curve, but rather its time derivative at the avoided
level crossings. In other words, the approximation of a linearly increasing
field $B$ around a certain transition point is sufficient to accurately
describe the dynamics at that transition.

\begin{figure}[tbp]
\includegraphics[width=8.8 cm]{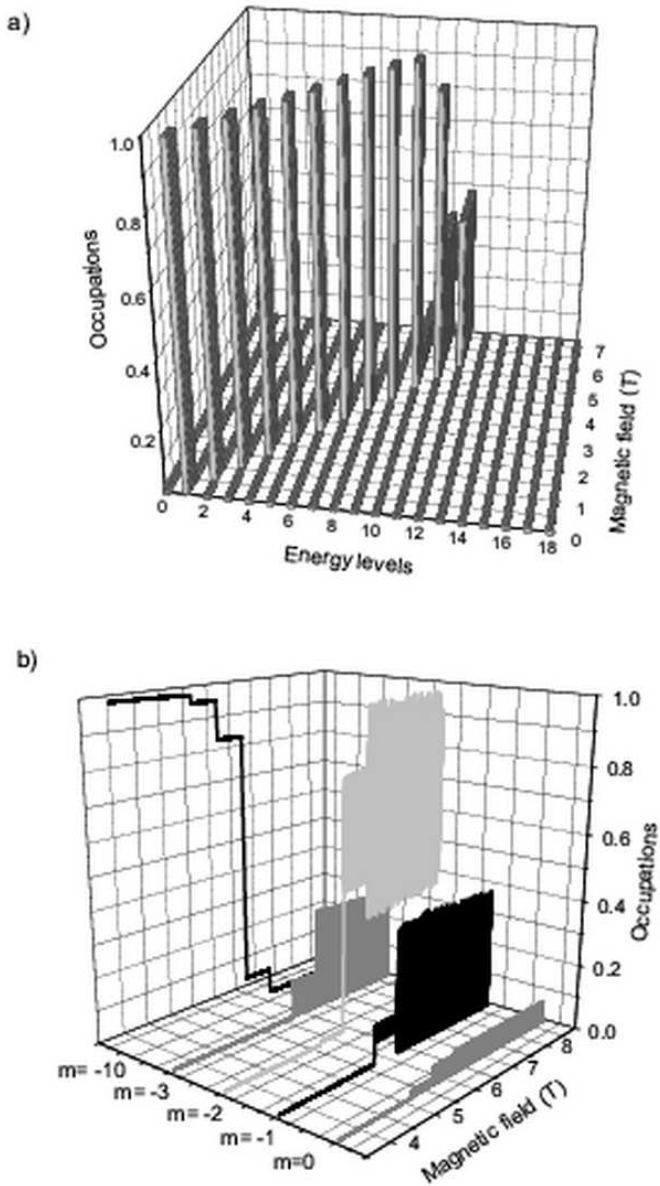}
\caption{/Reduced resolution for this preprint version./ The population of the levels corresponding to the states $\vert E_{n}
\rangle$ [a)] and $\vert m \rangle$ [b)] as a function of the external
magnetic field for a linear pulse with $w=1$ kT/s. For the sake of simplicity
in b) we show only populations larger than $0.05.$} \label{popfig}
\end{figure}

The population of the different eigenstates of $S_{z}$ and the adiabatic
states $\left| E_{n}(t)\right\rangle$ are shown in Fig.~\ref{popfig} as a
function of $B$ for the representative example of a pulse with linear shape
and $w=1$ kT/s. As we can see, at the beginning of the time evolution, when
the splitting of the adiabatic levels are very small and the magnetization is
almost constant, the population of the $|m=-10\rangle$ state does not change
significantly either. Around $B=5$ T the tunneling probability between
different states $|m\rangle$ and $|m^{\prime}\rangle$ becomes appreciable,
leading to the steps seen in Figs.~\ref{magnfig1} and \ref{magnfig2}. Note
that the population of the adiabatic levels show an opposite behavior:
initially practically all the population of the lower adiabatic level is
transferred to the higher one at the avoided crossings. On the other hand, for
larger external field values a nonzero population remains on the lower
adiabatic level, leading to a noticeable change of the magnetization. The
reason for the rapid oscillations in the populations of  $\vert m \rangle$
seen in Figs.~\ref{magnfig1}-\ref{popfig} is that at that high field strength
$S_{z}$ and $H_{S}$ do not commute, therefore the states $\vert m\rangle$ are
not exact eigenstates of the complete spin Hamiltonian. These oscillations
disappear when we include relaxation, see Sec.~\ref{relaxsec}.

If we restrict ourselves to the LZS model, then starting from the ground state
it is sufficient to find the first value of $B$ at which this adiabatic level
anticrosses the next one, calculate the LZS parameter $\Delta /\sqrt{\Omega}$
and use $P_{LZS}$ to obtain the population of the two relevant adiabatic
levels after the transition, and repeat this process until the end of the time
evolution. This approach includes the slow change of the levels $\left|
E_{n}(t)\right\rangle,$ causing a slight, continuous increase of the
magnetization as a function of $B$, but this effect is difficult to see in
Figs.~\ref{magnfig1} and \ref{magnfig2}: the steps dominate the behavior of
$\langle S_{z} \rangle.$ However, the LZS result obtained in this way is
quantitatively different from the exact $\langle S_{z}\rangle (t)$ curve that
was calculated by taking all the 21 levels into account (see
Fig.~\ref{magnfig1}). The position of the steps (determined by the avoided
crossings) are the same, but their heights are different, and this difference
can be as large as 30\%.

\begin{figure}[tbp]
\includegraphics[width=8.8 cm]{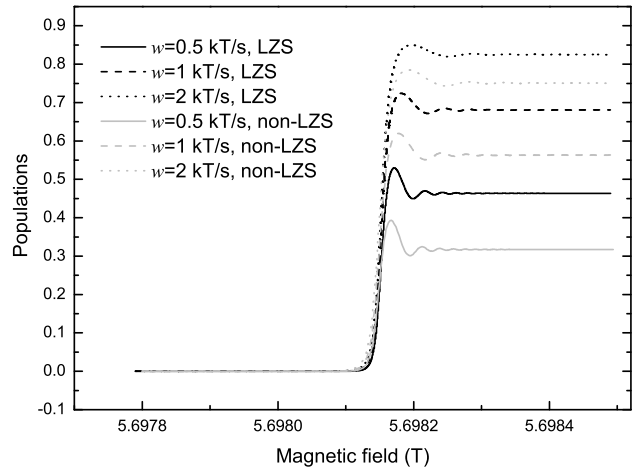}
\caption{The population of the level $E_{11}$ around the transition $\vert
E_{11} \rangle \rightarrow \vert E_{12} \rangle$ for different sweep rates. We
compare the exact results with those from the LZS model. The external field
pulse corresponding to this figure is given by Eq.~(\ref{pulse1}).}
\label{zoomfig}
\end{figure}

In order to understand the physical reason for this effect, we have to
investigate the relation between $P_{LZS}$ at a certain anticrossing and the
relevant transition probability resulting from the present calculation.
The first important point to take into account is that, to a very good
approximation, the transitions seen in Fig.~\ref{popfig} take place between
two neighboring adiabatic levels. For sweep rates in the kT/s range the
characteristic time of the transitions\cite{V99} at the avoided level
crossings neither overlap nor influence each other. Let us now concentrate on
a single step, as an example we focus on the vicinity of $B=5.7$ T (see
Fig.~\ref{zoomfig}), where the anticrossing levels $E_{11}$ and $E_{12}$
correspond to the states $\left\vert E_{11}\right\rangle \approx \left\vert
m=-10\right\rangle $ ($\left\vert m=-1\right\rangle$) and $\left\vert
E_{12}\right\rangle \approx \left\vert m=-1\right\rangle$ ($\left\vert
m=-10\right\rangle$) before (after) the transition.

The most remarkable point concerning Fig.~\ref{zoomfig} is that it was
obtained by assuming that \emph{only} the adiabatic levels $\left\vert
E_{11}\right\rangle $ and $ \left\vert E_{12}\right\rangle $ play a role in
the transition, the dynamics has been reduced to these two relevant levels.
Calculating the same transition by taking all the 21 levels into account shows
that this approximation estimates the exact dynamics with high accuracy. More
generally, it is always possible to perform a similar reduction around a given
avoided level crossing. Thus we obtain a method where a sequence of effective
two-level transitions can describe the time evolution. The results obtained in
this way are practically the same as those of the exact calculation, and
considering the numerical costs, this is a very effective method.

\begin{figure}[tbp]
\includegraphics[width=8.8 cm]{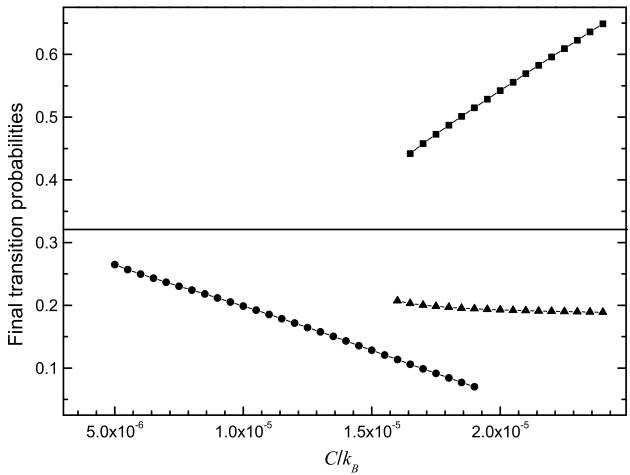}
\caption{The final population of the level $E_{11}$ after the transition
$\vert E_{11} \rangle \rightarrow \vert E_{12} \rangle$ around $B=5.7$ T  as a
function of the parameter $C$ in Eq.~(\ref{H1}). The level splitting
$\min(E_{12}- E_{11})$ is kept fixed, leading to the constant final population
in the LZS model indicated by the solid horizontal line. The magnetic field
pulse used is given by Eq.~(\ref{pulse1}) with $w=1$ kT/s.} \label{diffig}
\end{figure}

However, the time dependence of the expectation value of $ \langle
S_{z}\rangle $ obtained in this way is still different from the LZS result,
despite the fact that the LZS theory is also based on a two-level
approximation. Note that all the curves shown in Fig.~\ref{zoomfig} were
calculated using the same numerical method, the pulse shape ($f_{1}$), the
initial state and the level splitting were also the same for all the
calculated sweep rates: the only difference was the time dependence of the
energy levels and their coupling. (In fact, as it is clear from
Eqs.~(\ref{rhoexpand}), (\ref{neumann}) and (\ref{ine}), it is only the energy
difference $E_{12}(t)-E_{11}(t)$ that plays a role here.) These are multilevel
effects: the time dependence of the $2\times 2$ Hamiltonian obtained by the
reduction of $H_{S}$ to the relevant level pair is affected by all the other
levels, similarly to a renormalization effect. The influence of the states not
taking part in the transition results in a time dependence of the parameters
of the reduced Hamiltonian which is slightly different from the LZS model
described by Eq.~(\ref{Hred}). That is, the single parameter $\Delta/\sqrt{
\Omega }$ is not enough to describe these transitions, or in other words, we
have time dependent factors in the LZS matrix elements $\Delta (t), \Omega
(t).$ That is emphasized by calculating the long time limit transition
probability for the same transition shown in Fig.~\ref{zoomfig} with different
parameters in the Hamiltonian (\ref{H1}) in such a way, that the sweep rate
and the level splitting are fixed. As an example, we consider the level
splitting $\Delta$ at this transition as a function of only two parameters,
namely $C$ and $K$; we do not change $D,F$ and $E$ in $H_{S}.$ First we
calculate $\Delta(C,K)$ using the parameters given in Sec.~\ref{modelsec} and
obtain $\Delta_{0}=1.47\times 10^{-4}$ (in Kelvin). Then we solve the equation
$\Delta(C,K)=\Delta_{0}$ to find the parameter pairs $\{C,K\}$ that correspond
to the same level splitting as it was initially. It turns out that there are
several disconnected lines in the $C-K$ plane along which
$\Delta(C,K)=\Delta_{0}$, and consequently $P_{LZS}$ is constant for a given
sweep rate. But as we can see in Fig.~\ref{diffig}, the calculated non-LZS
transition probabilities, long after the transition took place, strongly
depend on the parameter $C$ in Eq.~(\ref{H1}): they range from 5\% up to 65\%,
and can be both higher and lower than $P_{LZS}$. Different symbols in
Fig.~\ref{diffig} correspond to different lines along which
$\Delta(C,K)=\Delta_{0}$. In any case, we can conclude that the final
transition probability at a given avoided crossing is also influenced by the
levels that do not take part in the relevant transition, and can therefore not
be described within the framework of the LZS model.

\section{Relaxation effects}
\label{relaxsec}

So far we considered unitary time evolution, i.e., the Hamiltonian (\ref{H})
governed the dynamics. However, there are interactions with e.g.~phonons that
are not included in $H_{S}.$ These additional degrees of freedom can be
considered as the environment of the spin system, and any realistic
description should take their influence into account.\cite{LL00,RL05,BFP05}
Additionally, as we shall see in this section, the rapid oscillations seen in
Figs.~\ref{magnfig1}-\ref{popfig} disappear on a very short time scale when
relaxation influences the dynamics.

Since phase relaxation is usually much faster than energy exchange between the
investigated quantum system and its environment, we concentrate on this kind
of decoherence, and assume a Lindblad-type\cite{L76} dynamical equation:
\begin{equation}
\frac{\partial \varrho }{\partial t}= -i\left[ H_{S}, \varrho %
\right]+\frac{\gamma}{2} \left( 2 S_{z} \varrho S_{z} -  S_{z}^{2} \varrho -
\varrho S_{z}^{2}\right) \label{master2}.
\end{equation}
The second term in this master equation will not change either the
magnetization of the sample nor the expectation value of $H_{0}$. It leads to
the gradual disappearance of the non-diagonal elements of $\varrho$ in the
eigenbasis of $S_{z}$ without changing the populations $\varrho_{n n}.$  The
result of this kind of relaxation is quite different during the transitions at
the avoided level crossings and between them. This difference is clearly seen
if we consider sweep rates of the order of kT/s, when the characteristic
transition times are $10^{-6}$--$10^{-7}$ s, while the spin system spends
about $10^{-3}$ s between two level crossings. These time scales should be
compared to $\gamma^{-1},$ which is in the range of $10^{-5}$--$10^{-7}$ s for
low temperatures ($T\approx2K$) at which several experiments were performed.
Thus, between two crossings, relaxation has enough time to destroy the phase
information almost completely.
\begin{figure}[tbp]
\includegraphics[width=8.8 cm]{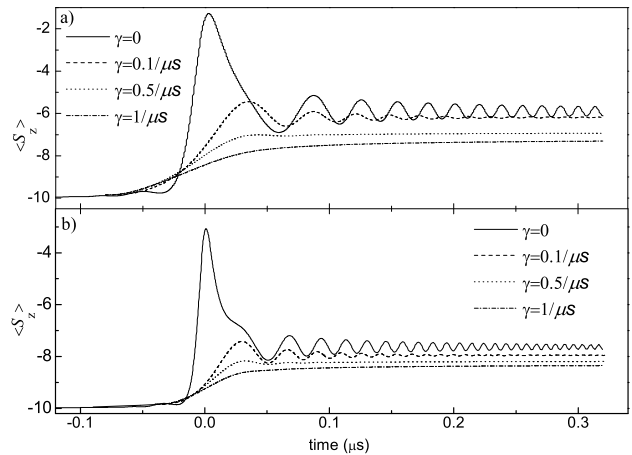}
\caption{The effect of phase relaxation on the dynamics of the magnetization
($\langle S_{z}\rangle$) around the transition $\vert E_{11} \rangle
\rightarrow \vert E_{12} \rangle.$ We assumed a linear pulse given by
Eq.~(\ref{pulse1}) with maximal sweep rate $w=1$ kT/s in part a) and $2$ kT/s
in part b). Note that the origin of the time axis has been shifted, $t=0$
corresponds to the time when $E_{12}-E_{11}$ is minimal. } \label{relaxfig}
\end{figure}

In other words, if after an avoided crossing the system is in a pure quantum
mechanical state $\vert \phi \rangle$, the initial condition at the next
crossing can be considered as the right hand side of the following scheme:
\begin{equation}
\left\vert \phi\right\rangle \left\langle \phi \right\vert=\sum_{n
m}\left\vert n\right\rangle \left\langle m \right\vert \varrho_{n m} \
\rightarrow \ \sum_{m}\left\vert m\right\rangle \left\langle m \right\vert
\varrho_{m m}. \label{dec}
\end{equation}
Note that the Hamiltonian part of the time evolution slightly modifies the
process above, but the final consequence, i.e., that dephasing changes the
initial conditions for the transitions at the avoided level crossings is still
valid. Decoherence as described by Eq.~(\ref{dec}) plays an important role in
the physical mechanism responsible for the observed form of hysteresis loops
(see Ref.~[\onlinecite{VG06}] for experimental results on systems of low-spin
magnetic molecules). We found that even for such fast external magnetic field
sweep rates as a few kT/s, quantum mechanical interference does not play an
important role in transitions at consecutive anticrossings, by the time the
system reaches the next transition point, the phase information has already
flown into the environment. For similar reasons the formation of closed
hysteresis curves cannot be influenced either by quantum interference.
However, even faster sweep rates, or faster return to a certain transition by
superimposing an oscillating magnetic field on a constant one around a certain
crossing, may give rise to quantum mechanical interference effects. We predict
that such interference phenomena will show up for magnetic field sweep rates
in the MT/s range. They will be seen as a very strong dependence of the form
of the hysteresis loops on the sweep rate due to the extreme sensitivity of
the process to the relative phases of the states that take part in a
transition.

The phenomena expected around transition regions are more interesting, as they
are consequences of the interplay between coherent effects and dephasing: the
characteristic time of the transitions are comparable, but usually shorter
than the dephasing time defined by $\gamma^{-1}.$ The general consequence of
the second term in the master equation (\ref{master2}) is that it decreases
the transition probability in the $\{|m\rangle \}$ basis, i.e., when
relaxation is present, it makes the steps in the magnetization smaller (see
Fig.~\ref{relaxfig}). The larger the value of $\gamma$, the stronger this
effects is, which underlines again the importance of the relative phase in
this physical system. In the previous section it was shown that phases gained
by the adiabatic states at an anticrossing can remarkably modify the
transition probability, now we see that loss in phase information has also
strong effects. Note that in contrast with the coherent case (see
Sec.~\ref{unitarysec}), the final transition probabilities in the presence of
dephasing do not scale with the sweep rate similarly to the LZS case. For
larger sweep rates the system spends less time in the transition region, and
consequently decoherence less strongly modifies the final transition
probability. That is, the larger the magnetic field sweep rate, the more
similar the dynamics around a transition becomes to the case without
dephasing. Additionally, we want to point out that the coherent oscillations,
the consequences of which has been seen in Figs.~\ref{magnfig1}-\ref{popfig},
are strongly damped even for weak dephasing.

However, the statement that the final transition probabilities are modified by
the presence of all the levels and thus cannot be accurately described within
the framework of a LZS model (even if we include relaxation as well) is also
true in the present case. Additionally, let us emphasize that the parameter
region discussed here is different from the strongly damped one studied in
Refs.~[\onlinecite{LL99,LL00}], where incoherent tunneling can give a proper
description. For external field sweep rates in the kT/s range, neither
coherent time evolution nor relaxation dominates, which leads to an
interesting interplay between these two qualitatively different processes.

\section{Conclusions}
We studied the time evolution of the spin degrees of freedom in molecular
nanomagnets, with a focus on the molecule Mn${}_{12}$-Ac in the presence of
time-dependent magnetic field. Using an appropriate 'exact numerical' method,
we followed the time evolution from zero external magnetic field until
saturation of the magnetization is reached. We found that for sweep rates in
the kT/s range, steps in the magnetization originate from two-level
transitions which cannot be described within the framework of the
Landau-Zener-St\"{u}ckelberg (LZS) model. This observation led us to the
introduction an efficient and accurate approximation based on two-level
non-LZS transitions. This method introduces the possibility of performing long
term dynamical calculations that can directly be related to experiments. We
also demonstrated that the sweep rate range of kT/s is special in the sense
that for realistic relaxation times, there are observable consequences of the
competition between coherent dynamics and decoherence that modify the heights
and widths of the magnetization steps.

\label{conclusionsec}

\acknowledgments We thank V.~V.~Moshchalkov and J.~Vanacken for the fruitful
discussions and O. K\'{a}lm\'{a}n for her valuable comments. This work was
supported by the Flemish-Hungarian Bilateral Programme, the Brazilian Council
for Research (CNPq), the Flemish Science Foundation (FWO-Vl), the Belgian
Science Policy and the Hungarian Scientific Research Fund (OTKA) under
Contracts Nos. T48888, D46043, M36803, M045596.


\begin{thebibliography}{35}
\expandafter\ifx\csname natexlab\endcsname\relax\def\natexlab#1{#1}\fi
\expandafter\ifx\csname bibnamefont\endcsname\relax
  \def\bibnamefont#1{#1}\fi
\expandafter\ifx\csname bibfnamefont\endcsname\relax
  \def\bibfnamefont#1{#1}\fi
\expandafter\ifx\csname citenamefont\endcsname\relax
  \def\citenamefont#1{#1}\fi
\expandafter\ifx\csname url\endcsname\relax
  \def\url#1{\texttt{#1}}\fi
\expandafter\ifx\csname urlprefix\endcsname\relax\def\urlprefix{URL }\fi
\providecommand{\bibinfo}[2]{#2} \providecommand{\eprint}[2][]{\url{#2}}

\bibitem[{\citenamefont{Gatteschi et~al.}(2006)\citenamefont{Gatteschi,
  Sessoli, and Villain}}]{GSV06}
\bibinfo{author}{\bibfnamefont{D.}~\bibnamefont{Gatteschi}},
  \bibinfo{author}{\bibfnamefont{R.}~\bibnamefont{Sessoli}}, \bibnamefont{and}
  \bibinfo{author}{\bibfnamefont{J.}~\bibnamefont{Villain}},
  \emph{\bibinfo{title}{Molecular Nanomagnets}} (\bibinfo{publisher}{Oxford
  University Press}, \bibinfo{year}{2006}).

\bibitem[{\citenamefont{Friedman et~al.}(1996)\citenamefont{Friedman, Sarachik,
  Tejada, and Ziolo}}]{FST96}
\bibinfo{author}{\bibfnamefont{J.~R.} \bibnamefont{Friedman}},
  \bibinfo{author}{\bibfnamefont{M.~P.} \bibnamefont{Sarachik}},
  \bibinfo{author}{\bibfnamefont{J.}~\bibnamefont{Tejada}}, \bibnamefont{and}
  \bibinfo{author}{\bibfnamefont{R.}~\bibnamefont{Ziolo}},
  \bibinfo{journal}{Phys. Rev. Lett.} \textbf{\bibinfo{volume}{76}},
  \bibinfo{pages}{3830} (\bibinfo{year}{1996}).

\bibitem[{\citenamefont{Mertes et~al.}(2001)\citenamefont{Mertes, Suzuki,
  Sarachik, Paltiel, Shtrikman, Zeldov, Rumberger, Hendrickson, and
  Christou}}]{MSS01}
\bibinfo{author}{\bibfnamefont{K.~M.} \bibnamefont{Mertes}},
  \bibinfo{author}{\bibfnamefont{Y.}~\bibnamefont{Suzuki}},
  \bibinfo{author}{\bibfnamefont{M.~P.} \bibnamefont{Sarachik}},
  \bibinfo{author}{\bibfnamefont{Y.}~\bibnamefont{Paltiel}},
  \bibinfo{author}{\bibfnamefont{H.}~\bibnamefont{Shtrikman}},
  \bibinfo{author}{\bibfnamefont{E.}~\bibnamefont{Zeldov}},
  \bibinfo{author}{\bibfnamefont{E.}~\bibnamefont{Rumberger}},
  \bibinfo{author}{\bibfnamefont{D.~N.} \bibnamefont{Hendrickson}},
  \bibnamefont{and} \bibinfo{author}{\bibfnamefont{G.}~\bibnamefont{Christou}},
  \bibinfo{journal}{Phys. Rev. Lett.} \textbf{\bibinfo{volume}{87}},
  \bibinfo{pages}{227205} (\bibinfo{year}{2001}).

\bibitem[{\citenamefont{Wernsdorfer et~al.}(2006)\citenamefont{Wernsdorfer,
  Murugesu, and Christou}}]{WMC06}
\bibinfo{author}{\bibfnamefont{W.}~\bibnamefont{Wernsdorfer}},
  \bibinfo{author}{\bibfnamefont{M.}~\bibnamefont{Murugesu}}, \bibnamefont{and}
  \bibinfo{author}{\bibfnamefont{G.}~\bibnamefont{Christou}},
  \bibinfo{journal}{Phys. Rev. Lett.} \textbf{\bibinfo{volume}{96}},
  \bibinfo{pages}{057208} (\bibinfo{year}{2006}).

\bibitem[{\citenamefont{Tejada et~al.}(2004)\citenamefont{Tejada, Chudnovsky,
  Hernandez, and Amig\'{o}}}]{TCHA04}
\bibinfo{author}{\bibfnamefont{J.}~\bibnamefont{Tejada}},
  \bibinfo{author}{\bibfnamefont{E.~M.} \bibnamefont{Chudnovsky}},
  \bibinfo{author}{\bibfnamefont{J.~M.} \bibnamefont{Hernandez}},
  \bibnamefont{and}
  \bibinfo{author}{\bibfnamefont{R.}~\bibnamefont{Amig\'{o}}},
  \bibinfo{journal}{Appl. Phys. Lett.} \textbf{\bibinfo{volume}{84}},
  \bibinfo{pages}{2373} (\bibinfo{year}{2004}).

\bibitem[{\citenamefont{Vanacken et~al.}(2004)\citenamefont{Vanacken,
  Stroobants, Malfait, Moshchalkov, Jordi, Tejada, Amigo, Chudnovsky, and
  Garanin}}]{VSM04}
\bibinfo{author}{\bibfnamefont{J.}~\bibnamefont{Vanacken}},
  \bibinfo{author}{\bibfnamefont{S.}~\bibnamefont{Stroobants}},
  \bibinfo{author}{\bibfnamefont{M.}~\bibnamefont{Malfait}},
  \bibinfo{author}{\bibfnamefont{V.~V.} \bibnamefont{Moshchalkov}},
  \bibinfo{author}{\bibfnamefont{M.}~\bibnamefont{Jordi}},
  \bibinfo{author}{\bibfnamefont{J.}~\bibnamefont{Tejada}},
  \bibinfo{author}{\bibfnamefont{R.}~\bibnamefont{Amigo}},
  \bibinfo{author}{\bibfnamefont{E.~M.} \bibnamefont{Chudnovsky}},
  \bibnamefont{and} \bibinfo{author}{\bibfnamefont{D.~A.}
  \bibnamefont{Garanin}}, \bibinfo{journal}{Phys. Rev. B}
  \textbf{\bibinfo{volume}{70}}, \bibinfo{pages}{220401}
  (\bibinfo{year}{2004}).

\bibitem[{\citenamefont{Hernandez-Minguez
  et~al.}(2005)\citenamefont{Hernandez-Minguez, Jordi, Amigo, Garcia-Santiago,
  Hernandez, and Tejada}}]{HJAGHT05}
\bibinfo{author}{\bibfnamefont{A.}~\bibnamefont{Hernandez-Minguez}},
  \bibinfo{author}{\bibfnamefont{M.}~\bibnamefont{Jordi}},
  \bibinfo{author}{\bibfnamefont{R.}~\bibnamefont{Amigo}},
  \bibinfo{author}{\bibfnamefont{A.}~\bibnamefont{Garcia-Santiago}},
  \bibinfo{author}{\bibfnamefont{J.~M.} \bibnamefont{Hernandez}},
  \bibnamefont{and} \bibinfo{author}{\bibfnamefont{J.}~\bibnamefont{Tejada}},
  \bibinfo{journal}{Europhys. Lett.} \textbf{\bibinfo{volume}{69}},
  \bibinfo{pages}{270} (\bibinfo{year}{2005}).

\bibitem[{\citenamefont{Dicke}(1954)}]{D54}
\bibinfo{author}{\bibfnamefont{R.~M.} \bibnamefont{Dicke}},
  \bibinfo{journal}{Phys. Rev.} \textbf{\bibinfo{volume}{93}},
  \bibinfo{pages}{439} (\bibinfo{year}{1954}).

\bibitem[{\citenamefont{Benedict et~al.}(1996)\citenamefont{Benedict, Ermolaev,
  Malyshev, Sokolov, and Trifonov}}]{SRAD}
\bibinfo{author}{\bibfnamefont{M.~G.} \bibnamefont{Benedict}},
  \bibinfo{author}{\bibfnamefont{A.~M.} \bibnamefont{Ermolaev}},
  \bibinfo{author}{\bibfnamefont{V.~A.} \bibnamefont{Malyshev}},
  \bibinfo{author}{\bibfnamefont{I.~V.} \bibnamefont{Sokolov}},
  \bibnamefont{and} \bibinfo{author}{\bibfnamefont{E.~D.}
  \bibnamefont{Trifonov}}, \emph{\bibinfo{title}{Superradiance}}
  (\bibinfo{publisher}{IOP}, \bibinfo{address}{Bristol}, \bibinfo{year}{1996}).

\bibitem[{\citenamefont{Chudnovsky and Garanin}(2002)}]{CG02}
\bibinfo{author}{\bibfnamefont{E.~M.} \bibnamefont{Chudnovsky}}
  \bibnamefont{and} \bibinfo{author}{\bibfnamefont{D.~A.}
  \bibnamefont{Garanin}}, \bibinfo{journal}{Phys. Rev. Lett.}
  \textbf{\bibinfo{volume}{89}}, \bibinfo{pages}{157201}
  (\bibinfo{year}{2002}).

\bibitem[{\citenamefont{Henner and Kaganov}(2003)}]{HK03}
\bibinfo{author}{\bibfnamefont{V.~K.} \bibnamefont{Henner}} \bibnamefont{and}
  \bibinfo{author}{\bibfnamefont{I.~V.} \bibnamefont{Kaganov}},
  \bibinfo{journal}{Phys. Rev. B} \textbf{\bibinfo{volume}{68}},
  \bibinfo{pages}{144420} (\bibinfo{year}{2003}).

\bibitem[{\citenamefont{Joseph et~al.}(2004)\citenamefont{Joseph, Calero, and
  Chudnovsky}}]{JCC04}
\bibinfo{author}{\bibfnamefont{C.~L.} \bibnamefont{Joseph}},
  \bibinfo{author}{\bibfnamefont{C.}~\bibnamefont{Calero}}, \bibnamefont{and}
  \bibinfo{author}{\bibfnamefont{E.~M.} \bibnamefont{Chudnovsky}},
  \bibinfo{journal}{Phys. Rev. B} \textbf{\bibinfo{volume}{70}},
  \bibinfo{pages}{174416} (\bibinfo{year}{2004}).

\bibitem[{\citenamefont{Benedict et~al.}(2005)\citenamefont{Benedict,
  F\"{o}ldi, and Peeters}}]{BFP05}
\bibinfo{author}{\bibfnamefont{M.~G.} \bibnamefont{Benedict}},
  \bibinfo{author}{\bibfnamefont{P.}~\bibnamefont{F\"{o}ldi}},
  \bibnamefont{and} \bibinfo{author}{\bibfnamefont{F.~M.}
  \bibnamefont{Peeters}}, \bibinfo{journal}{Phys. Rev. B}
  \textbf{\bibinfo{volume}{72}}, \bibinfo{pages}{214430}
  (\bibinfo{year}{2005}).

\bibitem[{\citenamefont{Suzuki et~al.}(2005)\citenamefont{Suzuki, Sarachik,
  Chudnovsky, McHugh, Gonzalez-Rubio, Avraham, Myasoedov, Zeldov, Shtrikman,
  Chakov et~al.}}]{SSC05}
\bibinfo{author}{\bibfnamefont{Y.}~\bibnamefont{Suzuki}},
  \bibinfo{author}{\bibfnamefont{M.~P.} \bibnamefont{Sarachik}},
  \bibinfo{author}{\bibfnamefont{E.~M.} \bibnamefont{Chudnovsky}},
  \bibinfo{author}{\bibfnamefont{S.}~\bibnamefont{McHugh}},
  \bibinfo{author}{\bibfnamefont{R.}~\bibnamefont{Gonzalez-Rubio}},
  \bibinfo{author}{\bibfnamefont{N.}~\bibnamefont{Avraham}},
  \bibinfo{author}{\bibfnamefont{Y.}~\bibnamefont{Myasoedov}},
  \bibinfo{author}{\bibfnamefont{E.}~\bibnamefont{Zeldov}},
  \bibinfo{author}{\bibfnamefont{H.}~\bibnamefont{Shtrikman}},
  \bibinfo{author}{\bibfnamefont{N.~E.} \bibnamefont{Chakov}},
  \bibnamefont{et~al.}, \bibinfo{journal}{Phys. Rev. Lett.}
  \textbf{\bibinfo{volume}{95}}, \bibinfo{pages}{147201}
  (\bibinfo{year}{2005}).

\bibitem[{\citenamefont{Hernandez-M\'{\i}nguez
  et~al.}(2005)\citenamefont{Hernandez-M\'{\i}nguez, Hernandez, Macia,
  Garc\'{\i}a-Santiago, Tejada, and Santos}}]{HMH05}
\bibinfo{author}{\bibfnamefont{A.}~\bibnamefont{Hernandez-M\'{\i}nguez}},
  \bibinfo{author}{\bibfnamefont{J.~M.} \bibnamefont{Hernandez}},
  \bibinfo{author}{\bibfnamefont{F.}~\bibnamefont{Macia}},
  \bibinfo{author}{\bibfnamefont{A.}~\bibnamefont{Garc\'{\i}a-Santiago}},
  \bibinfo{author}{\bibfnamefont{J.}~\bibnamefont{Tejada}}, \bibnamefont{and}
  \bibinfo{author}{\bibfnamefont{P.~V.} \bibnamefont{Santos}},
  \bibinfo{journal}{Phys. Rev. Lett.} \textbf{\bibinfo{volume}{95}},
  \bibinfo{pages}{217205} (\bibinfo{year}{2005}).

\bibitem[{\citenamefont{Leuenberger and Loss}(2001)}]{LL01}
\bibinfo{author}{\bibfnamefont{M.~N.} \bibnamefont{Leuenberger}}
  \bibnamefont{and} \bibinfo{author}{\bibfnamefont{D.}~\bibnamefont{Loss}},
  \bibinfo{journal}{Nature (London)} \textbf{\bibinfo{volume}{410}},
  \bibinfo{pages}{789} (\bibinfo{year}{2001}).

\bibitem[{\citenamefont{Landau}(1932)}]{L32}
\bibinfo{author}{\bibfnamefont{L.~D.} \bibnamefont{Landau}},
  \bibinfo{journal}{Phys. Z. Sowjetunion} \textbf{\bibinfo{volume}{2}},
  \bibinfo{pages}{46} (\bibinfo{year}{1932}).

\bibitem[{\citenamefont{Zener}(1932)}]{Z32}
\bibinfo{author}{\bibfnamefont{C.}~\bibnamefont{Zener}},
  \bibinfo{journal}{Proc. Roy. Soc. London, Ser. A}
  \textbf{\bibinfo{volume}{137}}, \bibinfo{pages}{696} (\bibinfo{year}{1932}).

\bibitem[{\citenamefont{St\"{u}ckelberg}(1932)}]{S32}
\bibinfo{author}{\bibfnamefont{E.~C.~G.} \bibnamefont{St\"{u}ckelberg}},
  \bibinfo{journal}{Helv. Phys. Acta} \textbf{\bibinfo{volume}{5}},
  \bibinfo{pages}{369} (\bibinfo{year}{1932}).

\bibitem[{\citenamefont{Mirebeau et~al.}(1999)\citenamefont{Mirebeau, Hennion,
  Casalta, Andres, G\"{u}del, Irodova, and Caneschi}}]{M99}
\bibinfo{author}{\bibfnamefont{I.}~\bibnamefont{Mirebeau}},
  \bibinfo{author}{\bibfnamefont{M.}~\bibnamefont{Hennion}},
  \bibinfo{author}{\bibfnamefont{H.}~\bibnamefont{Casalta}},
  \bibinfo{author}{\bibfnamefont{H.}~\bibnamefont{Andres}},
  \bibinfo{author}{\bibfnamefont{H.~U.} \bibnamefont{G\"{u}del}},
  \bibinfo{author}{\bibfnamefont{A.~V.} \bibnamefont{Irodova}},
  \bibnamefont{and} \bibinfo{author}{\bibfnamefont{A.}~\bibnamefont{Caneschi}},
  \bibinfo{journal}{Phys. Rev. Lett.} \textbf{\bibinfo{volume}{83}},
  \bibinfo{pages}{628} (\bibinfo{year}{1999}).

\bibitem[{\citenamefont{Barra et~al.}(1997)\citenamefont{Barra, Gatteschi, and
  Sessoli}}]{BGS97}
\bibinfo{author}{\bibfnamefont{A.~L.} \bibnamefont{Barra}},
  \bibinfo{author}{\bibfnamefont{D.}~\bibnamefont{Gatteschi}},
  \bibnamefont{and} \bibinfo{author}{\bibfnamefont{R.}~\bibnamefont{Sessoli}},
  \bibinfo{journal}{Phys. Rev. B} \textbf{\bibinfo{volume}{56}},
  \bibinfo{pages}{8192} (\bibinfo{year}{1997}).

\bibitem[{\citenamefont{Hill et~al.}(1998)\citenamefont{Hill, Perenboom, Dalal,
  Hathaway, Stalcup, and Brooks}}]{H98}
\bibinfo{author}{\bibfnamefont{S.}~\bibnamefont{Hill}},
  \bibinfo{author}{\bibfnamefont{J.~A. A.~J.} \bibnamefont{Perenboom}},
  \bibinfo{author}{\bibfnamefont{N.~S.} \bibnamefont{Dalal}},
  \bibinfo{author}{\bibfnamefont{T.}~\bibnamefont{Hathaway}},
  \bibinfo{author}{\bibfnamefont{T.}~\bibnamefont{Stalcup}}, \bibnamefont{and}
  \bibinfo{author}{\bibfnamefont{J.~S.} \bibnamefont{Brooks}},
  \bibinfo{journal}{Phys. Rev. Lett.} \textbf{\bibinfo{volume}{80}},
  \bibinfo{pages}{2453} (\bibinfo{year}{1998}).

\bibitem[{\citenamefont{Hill et~al.}(2003)\citenamefont{Hill, Edwards, Jones,
  Dalal, and North}}]{HEJ03}
\bibinfo{author}{\bibfnamefont{S.}~\bibnamefont{Hill}},
  \bibinfo{author}{\bibfnamefont{R.~S.} \bibnamefont{Edwards}},
  \bibinfo{author}{\bibfnamefont{S.~I.} \bibnamefont{Jones}},
  \bibinfo{author}{\bibfnamefont{N.~S.} \bibnamefont{Dalal}}, \bibnamefont{and}
  \bibinfo{author}{\bibfnamefont{J.~M.} \bibnamefont{North}},
  \bibinfo{journal}{Phys. Rev. Lett.} \textbf{\bibinfo{volume}{90}},
  \bibinfo{pages}{217204} (\bibinfo{year}{2003}).

\bibitem[{\citenamefont{van Vleck}(1929)}]{vV29}
\bibinfo{author}{\bibfnamefont{J.~H.} \bibnamefont{van Vleck}},
  \bibinfo{journal}{Phys. Rev.} \textbf{\bibinfo{volume}{33}},
  \bibinfo{pages}{467} (\bibinfo{year}{1929}).

\bibitem[{\citenamefont{des Cloizeaux}(1960)}]{dC60}
\bibinfo{author}{\bibfnamefont{J.}~\bibnamefont{des Cloizeaux}},
  \bibinfo{journal}{Nucl. Phys.} \textbf{\bibinfo{volume}{20}},
  \bibinfo{pages}{321} (\bibinfo{year}{1960}).

\bibitem[{\citenamefont{Klein}(1974)}]{K74}
\bibinfo{author}{\bibfnamefont{D.~J.} \bibnamefont{Klein}},
  \bibinfo{journal}{J. Chem. Phys.} \textbf{\bibinfo{volume}{61}},
  \bibinfo{pages}{786} (\bibinfo{year}{1974}).

\bibitem[{\citenamefont{Garanin}(1991)}]{G91}
\bibinfo{author}{\bibfnamefont{D.~A.} \bibnamefont{Garanin}},
  \bibinfo{journal}{J. Phys. A} \textbf{\bibinfo{volume}{24}},
  \bibinfo{pages}{L61} (\bibinfo{year}{1991}).

\bibitem[{\citenamefont{Leuenberger and Loss}(2000)}]{LL00}
\bibinfo{author}{\bibfnamefont{M.~N.} \bibnamefont{Leuenberger}}
  \bibnamefont{and} \bibinfo{author}{\bibfnamefont{D.}~\bibnamefont{Loss}},
  \bibinfo{journal}{Phys. Rev. B.} \textbf{\bibinfo{volume}{61}},
  \bibinfo{pages}{1286} (\bibinfo{year}{2000}).

\bibitem[{\citenamefont{Yoo and Park}(2005)}]{YP05}
\bibinfo{author}{\bibfnamefont{S.-K.} \bibnamefont{Yoo}} \bibnamefont{and}
  \bibinfo{author}{\bibfnamefont{C.-S.} \bibnamefont{Park}},
  \bibinfo{journal}{Phys. Rev. B} \textbf{\bibinfo{volume}{71}},
  \bibinfo{pages}{012409} (\bibinfo{year}{2005}).

\bibitem[{\citenamefont{Liu et~al.}(2002)\citenamefont{Liu, Wu, Fu, Diener, and
  Niu}}]{LWF02}
\bibinfo{author}{\bibfnamefont{J.}~\bibnamefont{Liu}},
  \bibinfo{author}{\bibfnamefont{B.}~\bibnamefont{Wu}},
  \bibinfo{author}{\bibfnamefont{L.}~\bibnamefont{Fu}},
  \bibinfo{author}{\bibfnamefont{R.~B.} \bibnamefont{Diener}},
  \bibnamefont{and} \bibinfo{author}{\bibfnamefont{Q.}~\bibnamefont{Niu}},
  \bibinfo{journal}{Phys. Rev. B} \textbf{\bibinfo{volume}{65}},
  \bibinfo{pages}{224401} (\bibinfo{year}{2002}).

\bibitem[{\citenamefont{Vitanov}(1999)}]{V99}
\bibinfo{author}{\bibfnamefont{N.~V.} \bibnamefont{Vitanov}},
  \bibinfo{journal}{Phys. Rev. A} \textbf{\bibinfo{volume}{59}},
  \bibinfo{pages}{988} (\bibinfo{year}{1999}).

\bibitem[{\citenamefont{Rousochatzakis and Luban}(2005)}]{RL05}
\bibinfo{author}{\bibfnamefont{I.}~\bibnamefont{Rousochatzakis}}
  \bibnamefont{and} \bibinfo{author}{\bibfnamefont{M.}~\bibnamefont{Luban}},
  \bibinfo{journal}{Phys. Rev. B} \textbf{\bibinfo{volume}{72}},
  \bibinfo{pages}{134424} (\bibinfo{year}{2005}).

\bibitem[{\citenamefont{Lindblad}(1976)}]{L76}
\bibinfo{author}{\bibfnamefont{G.}~\bibnamefont{Lindblad}},
  \bibinfo{journal}{Commun. math. Phys.} \textbf{\bibinfo{volume}{48}},
  \bibinfo{pages}{119} (\bibinfo{year}{1976}).

\bibitem[{\citenamefont{Vogelsberger and Garanin}(2006)}]{VG06}
\bibinfo{author}{\bibfnamefont{M.}~\bibnamefont{Vogelsberger}}
  \bibnamefont{and} \bibinfo{author}{\bibfnamefont{D.~A.}
  \bibnamefont{Garanin}}, \bibinfo{journal}{Phys. Rev. B}
  \textbf{\bibinfo{volume}{73}}, \bibinfo{pages}{092412}
  (\bibinfo{year}{2006}).

\bibitem[{\citenamefont{Leuenberger and Loss}(1999)}]{LL99}
\bibinfo{author}{\bibfnamefont{M.~N.} \bibnamefont{Leuenberger}}
  \bibnamefont{and} \bibinfo{author}{\bibfnamefont{D.}~\bibnamefont{Loss}},
  \bibinfo{journal}{Europhys. Lett.} \textbf{\bibinfo{volume}{45}},
  \bibinfo{pages}{692} (\bibinfo{year}{1999}).

\end{thebibliography}
\end{document}